\documentclass{appolb}
\usepackage{epsfig}
\usepackage{amsmath}

\begin{document}
\title{Chiral Symmetry Breaking, Trace Anomaly and Baryons\\
in Hot and Dense Matter%
\thanks{Presented at the HIC for FAIR Workshop and XXVIII Max Born Symposium,
Wroclaw, Poland, May 19-21 2011.}%
}
\author{Chihiro Sasaki
\address{Frankfurt Institute for Advanced Studies,
D-60438 Frankfurt am Main,
Germany}
}
\maketitle
\begin{abstract}
We propose an effective chiral Lagrangian with a chiral scalar introduced
as a dilaton associated with broken conformal symmetry and responsible for 
the trace anomaly in QCD and discuss the properties of hadronic matter at high 
density and temperature.
As the ``dilaton limit'' is taken, which drives a system from nuclear matter
density to near chiral restoration density, a linear sigma model emerges from
the highly non-linear structure. 
A striking prediction is that as the dilaton limit is approached,
the omega-nucleon interaction gets strongly suppressed at high density.
This is shown to be a firm statement at the quantum level protected by 
an infrared fixed point of the renormalization group equations derived
in chiral perturbation theory.
\end{abstract}
\PACS{21.30.Fe, 12.39.Fe, 21.65.Mn}

\section{Introduction}

Non-perturbative aspects of QCD in low energies are expected to be
captured using effective field theories constructed based on global symmetries 
of QCD Lagrangian and their breaking pattern. In the limit of massless quarks 
the Lagrangian possesses the chiral symmetry and scale invariance, both
of which are dynamically broken in the physical vacuum due to the strong 
interaction. The QCD trace anomaly signals the emergence of a scale at the
quantum level from the theory without any dimension-full parameters~\cite{trace}.
Thus spontaneous chiral symmetry breaking, which gives rise to a nucleon
mass, and the trace anomaly are closely linked to each other~\cite{bardeen} and 
dynamical scales in hadronic systems are considered to originate from them.
How they behave under extreme conditions such as high temperature and density
is one of the main issues in QCD~\cite{review}.

In nuclear physics, a scalar meson plays an essential role
as known from Walecka model that works fairly well for phenomena near nuclear
matter density~\cite{walecka}. On the other hand, at high density, the relevant 
Lagrangian that has correct symmetry is the linear sigma model, and the scalar 
needed there is the fourth component of the chiral four-vector $(\vec{\pi},\sigma)$.
Thus in order to probe highly hot/dense matter, we have to figure out how the
chiral scalar at low temperature/density transmutes to the fourth component
of the four-vector.
In this contribution, we construct an effective chiral Lagrangian for hadrons 
implemented by the conformal invariance introducing a chiral scalar as a dilaton 
associated with broken conformal symmetry and responsible for the trace anomaly 
in QCD. As the ``dilaton limit''~\cite{vanKolck} is taken, which drives a system 
from nuclear matter density to near chiral restoration density, a linear sigma 
model emerges from the highly non-linear structure with the omega meson decoupling 
from the nucleons~\cite{SLPR}. We also show a conceivable link of the dilaton
limit at quantum level to an infrared fixed point of the renormalization group 
equations formulated in chiral perturbation theory with the lowest-lying 
parity-doubled nucleons~\cite{PLRS}.

\section{Role of Dilatons near Chiral Symmetry Restoration}

The trace anomaly is implemented in a chiral Lagrangian by introducing
a dilaton (or glueball) field representing the gluon condensate 
$\langle G_{\mu\nu}G^{\mu\nu} \rangle$~\cite{schechter}. 
Following~\cite{miransky}, we write the trace anomaly in terms of ``soft'' 
dilaton $\chi_s$ and ``hard'' dilaton $\chi_h$. The dilaton potential,
\begin{equation}
V(\chi) = V_s(\chi_s) + V_h(\chi_h)\,,
\end{equation}
is assumed to have a negligible mixing between soft and hard sectors
in order to avoid an undesirably strong coupling of the glueball to pions.
As suggested in~\cite{LeeRho}, 
we will associate the soft dilaton with that component locked to the quark 
condensate $\langle\bar{q}q\rangle$. We assume that this is the component which 
``melts'' across the chiral phase transition whereas the hard component remaining 
non-vanishing~\footnote{
  The ``melting'' of the soft component is observed in 
  dynamical lattice calculation in temperature~\cite{miller} but is 
  an assumption in density.
}.
It was shown in~\cite{LeeRho} that the soft dilaton
plays an important role in the emergence of a half-skyrmion phase at high
density where a skyrmion turns into two half-skyrmions~\cite{half}.

In introducing baryonic degrees of freedom, there are two alternative ways of 
assigning chirality to the nucleons. One is the ``naive'' assignment~\footnote{
  We put this terminology in a quotation mark since it is a misnomer, used merely 
  to distinguish it from the alternative option.
} and the other the mirror assignment. The ``naive'' assignment,
\begin{equation}
\psi_L \to L\psi_L\,,
\quad
\psi_R \to R\psi_R\,,
\end{equation}
is anchored on 
the standard chiral symmetry structure where the entire constituent quark or 
nucleon mass (in the chiral limit) is generated by spontaneous symmetry breaking.
The alternative, mirror assignment~\cite{dk,mirror}, 
\begin{eqnarray}
&&
\psi_{1L} \to L\psi_{1L}\,,
\quad
\psi_{1R} \to R\psi_{1R}\,,
\nonumber\\
&&
\psi_{2L} \to R\psi_{2L}\,,
\quad
\psi_{2R} \to L\psi_{2R}\,,
\end{eqnarray}
allows a chiral invariant mass term,
\begin{equation}
{\mathcal L}_m
= m_0\left( \bar{\psi}_2\gamma_5\psi_1 - \bar{\psi}_1\gamma_5\psi_2\right)\,,
\end{equation} 
which remains non-zero at chiral restoration.
This means that a part of the nucleon mass, $m_0$, must arise 
from a mechanism that is not associated with spontaneous chiral symmetry breaking. 
At present, analysis of various observables both in the 
vacuum such as pion-nucleon scattering etc. and in medium such as nuclear matter 
properties etc. based on linear and nonlinear sigma models with mirror 
baryons~\cite{nemoto,analysis} cannot rule out an $m_0$ of a few hundred MeV. 
As one approaches the chiral restoration point, the two
assignments, even if indistinguishable at low density/temperature,
are expected to start showing their differences.
The origin of such a mass $m_0$ can be traced back to the non-vanishing
gluon condensate in chiral symmetric phase and therefore the broken scale 
symmetry is possessed by the hard dilaton.
In this way we attribute the origin of $m_0$ to the hard component of the gluon 
condensate, which is chiral invariant~\cite{SLPR}.

What about the low-lying meson masses? Mended symmetry is the algebraic
consequence of spontaneously broken chiral symmetry along with the assumption
on the scattering amplitudes of Nambu-Goldstone bosons in large $N_c$ and the 
mesons are assembled into a few of the irreducible representations~\cite{weinberg}.
A prediction of the mended symmetry near chiral symmetry restoration is that
the pion and other lowest-lying mesons become massless and fill out
a full representation of the chiral group. Generally, a chirally-invariant 
mass for the mesons, as introduced for the parity-doubled nucleons, is not 
excluded.

\section{Dilaton Limit}

Our aim is to derive an effective Lagrangian for the Nambu-Goldstone bosons,
vector mesons and soft dilatons in the linear basis starting with the hidden 
local symmetric (HLS) Lagrangian following the strategy of Beane and van Kolck.
The 2-flavored HLS Lagrangian is based on 
a $G_{\rm{global}} \times H_{\rm{local}}$ symmetry,
where $G_{\rm global}=[SU(2)_L \times SU(2)_R]_{\rm global}$
is the chiral symmetry and $H_{\rm local}=[SU(2)_V]_{\rm local}$
is the HLS~\cite{hls}.
The entire symmetry $G_{\rm global}\times H_{\rm local}$
is spontaneously broken to a diagonal $SU(2)_V$.
The basic quantities are the HLS gauge boson, $V_\mu$,
and two matrix valued variables $\xi_L$, $\xi_R$,
which are combined in a $2 \times 2$ special-unitary matrix
$U = \xi_L^\dagger \xi_R$.
Conformal invariance can be embedded in chiral Lagrangians by
introducing a scalar field $\tilde{\chi}$ via $\chi = F_\chi\tilde{\chi}$
and $\kappa = (F_\pi/F_\chi)^2$~\cite{vanKolck}.

\begin{figure}
\begin{center}
\includegraphics[width=7cm]{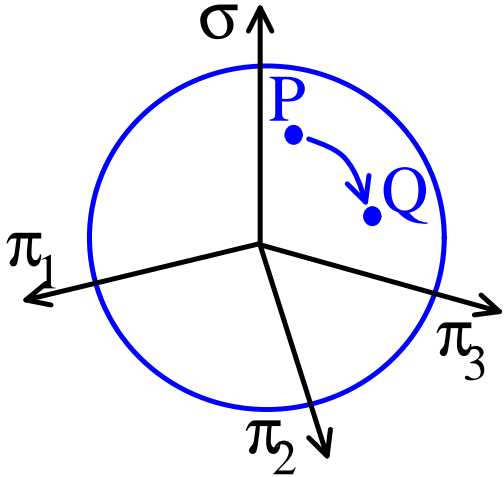}
\caption{
A chiral sphere in $(\vec{\pi},\sigma)$ space. 
$P$ on the sphere is mapped to another point $Q$ 
via chiral transformations.
}
\label{o4}
\end{center}
\end{figure}

Near chiral symmetry restoration the quarkonium component of the
dilaton field becomes a scalar mode which forms with pions an O(4)
quartet~\cite{vanKolck}. This can be formulated by making a transformation
of a non-linear chiral Lagrangian to a linear basis exploiting the dilaton limit. 
Let $\Phi$ be the basic building block of a linear sigma model 
$\Phi = \sigma + i\vec{\tau}\cdot\vec{\pi}$ which transforms as 
$\Phi \to L\Phi R^\dagger$. Through chiral transformations a point
on the 4-dimensional sphere is mapped to another point (see Fig.~\ref{o4}).
One can also express $\Phi$ in polar coordinates under the constraint
$F_\pi^2 = \sqrt{\sigma^2 + \vec{\pi}^2}$, where a point is specified
by three angles $\vec{\theta}=(\theta_1,\theta_2,\theta_3)$. Utilizing
the polar decomposition the linear sigma model Lagrangian is rewritten
to the standard non-linear chiral Lagrangian.
In the following we show a linearized Lagrangian assuming two different
chirality assignments to the positive and negative parity nucleons, 
the ``naive'' and mirror assignments.

In the ``naive'' model we introduce new fields, $\Sigma$ and ${\mathcal N}$, as
\begin{eqnarray}
\Sigma
&=& \xi_L^\dagger\xi_R \chi\sqrt{\kappa}
= s + i\vec{\tau}\cdot\vec{\pi}\,,
\\
{\mathcal N}
&=& \frac{1}{2}\left[ \left( \xi_R^\dagger + \xi_L^\dagger \right)
{}+ \gamma_5\left( \xi_R^\dagger - \xi_L^\dagger \right) \right] N\,.
\end{eqnarray}
The linearized Lagrangian includes terms which generate singularities,
negative powers of $\mbox{tr}\left[ \Sigma\Sigma^\dagger\right]$,
in chiral symmetric phase. Those terms carry the following factor:
\begin{equation}
X_N = g_V - g_A\,,
\quad
X_\chi = 1-\kappa\,.
\end{equation}
Assuming that nature disallows any singularities in the case considered, 
we require that they be absent in the Lagrangian, i.e.
$X_N = X_\chi = 0$. We find $\kappa = 1$ and $g_A = g_V$.
A particular value, $g_V = g_A = 1$, recovers the large $N_c$
algebraic sum rules~\cite{vanKolck}. Thus, we adopt the dilaton
limit as
\begin{equation}
\kappa = g_A = g_V = 1\,.
\end{equation}
The special value, $g_V=1$, is in fact achieved as a fixed point of the
renormalization group equations formulated in the chiral perturbation
theory with HLS when one approaches chiral restoration
from the low density or temperature side~\cite{PLRS}.

A noteworthy feature of the dilaton-limit Lagrangian is that the vector mesons 
decouple from the nucleons while their coupling to the Goldstone bosons remains.
This has two striking new predictions. Taking the dilaton limit drives the Yukawa 
interaction to vanish as $g_{VN}^2=(g\,(1-g_V))^2\rightarrow 0$ for 
$V=\rho, \omega$ for any finite value of the HLS gauge coupling $g$. In HLS for 
the meson sector, the model has the vector manifestation (VM) fixed point as one 
approaches chiral symmetry restoration, therefore the HLS coupling $g$ also tends 
to zero 
proportional to the quark condensate. It thus follows that combined with the VM, 
the coupling $g_{V N}$ will tend to vanish rapidly near the phase transition point. 
In nuclear forces, what is effective is the ratio $g_{VN}^2/m_V^2$ which goes 
as $(1-g_V)^2$. This means that (1) the two-body repulsion which holds two 
nucleons apart at short distance will be suppressed in dense medium and 
(2) the symmetry energy going as $S_{\rm sym}\propto g_{\rho N}^2$ will also 
get suppressed. As a principal consequence, the EoS at some high density 
approaching the dilaton limit will become softer {\it even without such exotic 
happenings as kaon condensation or strange quark matter}.

In the present scheme, the shortest-range component of the three-body forces 
also vanishes in the dilaton limit. The one-pion exchange three-body force 
involving a contact two-body force will also get suppressed as 
$\sim g_{\omega N}^2$. Thus only the longest-range two-pion exchange three-body 
forces will remain operative at large density in compact stars. How this 
intricate mechanism affects the EoS at high density is a challenging issue to 
resolve.

The dilaton limit is unchanged by the mirror baryons and therefore one
arrives at similar phenomenological consequences to those mentioned above.
The quenching of the short-range repulsion is independent of the chirality 
assignment of the nucleon and this is indicative of a universality of the 
short-distance interaction.
How large is $m_0$ at the chiral symmetry restoration? 
It seems natural to expect that the source for non-zero $m_0$ is in the hard 
dilaton condensate. 
A rough estimate can be made from thermodynamic considerations and the gluon
condensate calculated on a lattice in the presence of
dynamical quarks known to be~\cite{miller}
\begin{equation}
\langle G_{\mu\nu}G^{\mu\nu}\rangle_{T_{\rm ch}}
\simeq \frac{1}{2}\langle G_{\mu\nu}G^{\mu\nu}\rangle_{T=0}\,,
\end{equation}
at pseudo-critical temperature $T_{\rm ch} \sim 170$ MeV.
Adopting the bag constant and mass for the hard dilaton as
\begin{equation}
B_h(T_{\rm ch}) = \frac{1}{2}B(T=0)\,,
\quad
m_{\chi_h}^2 = \frac{1}{2}m_G^2\,,
\end{equation}
one finds $m_0 = 210$ MeV. This is in agreement with the estimate made in vacuum 
phenomenology~\cite{PLRS}. The nucleon in the mirror model stays massive
at chiral symmetry restoration, so a different EoS from that in the ``naive''
model would be expected. This issue and more realistic estimate of $m_0$
need to be carried out.

The axial-vector meson which figures in the mended symmetry can be dealt with on 
the same footing with the others and is introduced by generalizing 
$H_{\rm local}$ to $G_{\rm local}$ (GHLS) so that the entire symmetry of the
theory becomes $G_{\rm global} \times G_{\rm local}$~\cite{hls,GHLS}.
Applying the same procedure as before, the non-linear GHLS Lagrangian with
introducing a soft dilaton is transformed to its linearized form. One arrives
at the vector and axial-vector meson masses proportional to the chiral order 
parameter.
Thus, when chiral symmetry restoration takes place the mended symmetry becomes
manifest. Introducing a chiral invariant mass for the mesons will modify
the value of $m_0$.

Quantum loop corrections are systematically calculated in a chiral perturbation
theory (ChPT) with HLS. In the ``naive'' assignment we assign the chiral 
counting ${\mathcal O}(p)$ to the nucleon mass,
\begin{equation}
m_N \sim {\mathcal O}(p)\,,
\end{equation}
and the one-loop diagrams are evaluated in the relativistic formalism.
In the mirror assignment the nucleon mass is not entirely generated by
spontaneous chiral symmetry breaking and we identify the origin of
the chiral invariant mass $m_0$ with the explicit breaking of the QCD
scale invariance, i.e. a hard dilaton, which has no direct link  with the 
chiral dynamics. Consider $m_0$ to be large compared with dynamically generated 
mass~\footnote{
 We associate $m_0$ with the hard dilaton and thus the quantity $m_{N_\pm} - m_0$ 
 should {\it conceptually} be compatible to $\Lambda_{\rm QCD}$ even though $m_0$ 
 is a few hundred MeV as given above. This is changed e.g. in the 
 presence of a scalar tetraquark state to be around $500$ MeV~\cite{analysis}.
} 
and adopt a heavy baryon chiral perturbation theory (HBChPT)~\cite{HBChPT} 
in the presence of $m_0$. We write the nucleon momentum as
\begin{equation}
p^\mu = m_N v^\mu + k^\mu\,,
\end{equation}
where $v^\mu$ is the four-velocity with $v^2=1$ and $k^\mu$ is the
residual momentum of order $\Lambda_{\rm QCD}$, so that one can
perform a chiral perturbation theory systematically in energy range
below the chiral symmetry breaking scale $\Lambda_{\chi} \sim 1$ GeV.
A heavy-baryon doublet $B$ is defined by
\begin{equation}
\begin{pmatrix}
B_+
\\
B_-
\end{pmatrix}
=
\exp\left[ i m_0 v\cdot x \right]
\begin{pmatrix}
N_+
\\
N_-
\end{pmatrix}\,.
\end{equation}
Interestingly, the ChPT with either chirality assignment yields an infrared fixed
point of the coupled renormalization group equations which can be identified
with the dilaton limit~\cite{PLRS}.

\section{Conclusions and Remarks}

\begin{figure}
\begin{center}
\includegraphics[width=11cm]{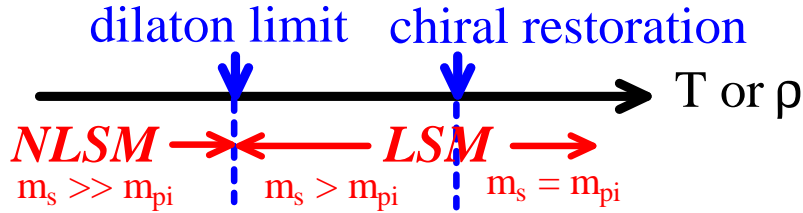}
\caption{
Expected changeover of effective theories near chiral symmetry
restoration.
}
\label{switch}
\end{center}
\end{figure}

We have shown how an effective theory near chiral symmetry restoration
emerges from the dilaton-implemented HLS Lagrangian at the dilaton limit
as illustrated in Fig.~\ref{switch},
and discussed its phenomenological implications at high baryon density.
The soft dilaton is responsible for the spontaneous breaking of the scale
symmetry and its condensate vanishes when the chiral symmetry is restored.
In fact, topological stability of the half-skyrmion phase has been 
observed~\cite{half}. This is a strong indication that the configuration is
robust and it could be associated with the scale symmetry restoration at
high density in continuum theories.

One important prediction is that the repulsion at short distance in nuclear 
interactions should get suppressed at a density in the vicinity of the dilaton 
limit. Another hitherto unsuspected result is that the symmetry energy which 
plays a crucial role in the structure of compact stars also should get suppressed. 
Put together, they will soften the EoS of compact-star matter at some high density.
An interesting possibility is that our mechanism could accommodate an exotica-free 
nucleon-only EoS (such as AP4 in Fig. 3 of Ref.~\cite{ozel}) with a requisite 
softening at higher density that could be compatible with the 
$1.97 \pm 0.04\,M_\odot$ neutron star data~\cite{2solarmass}.
It is an interesting and feasible phenomenological application of this model
to determine the EoS and in-medium condensate of the dilaton as well
as the onset of the dilaton limit at high density under a certain, e.g. mean field,
approximation.

Nuclear structure studies tell us that the ``hard core'' is not a physical 
observable in medium, that is, it is not visible but shoved under what is 
known as ``short-range correlation''. In fact, nuclear structure approaches 
anchored on effective field theory and renormalization group show that the 
``hard-core'' repulsion present in two-nucleon potentials plays no role in 
low-energy physical observables~\cite{kuo}.
Within the field theoretical framework we are working with, the short-distance 
repulsion is suppressed in the background or ``vacuum'' defined by density and
the mended symmetry point of view would offer a possible way to understand it.

Our main observation on the suppressed repulsive interaction is a common
feature in the two different assignments, ``naive'' and mirror, of chirality.
Furthermore, the dilaton limit turns out to be an IR fixed-point
of the renormalization group equations formulated in the chiral perturbation
theory with HLS. Therefore decoupling of vector mesons from
nucleons is a firm statement at quantum level. The derivative expansion in the
HLS theory is justified for small gauge coupling, $g \sim {\mathcal O}(p)$,
and in the limit of $g \to 0$ the symmetry of the Lagrangian is in fact
enlarged, which is known as ``vector realization'' of Georgi~\cite{georgi}
and could protect the dilaton limit at quantum level.
The nucleon mass near chiral symmetry restoration exhibits a striking
difference in the two scenarios. 
How the dilaton-limit suppression of the repulsion -- which seems to be universal 
independent of the assignments but may manifest itself differently in the two 
cases -- will affect the EoS for compact stars is an interesting question to 
investigate.

In the scalar sector of low-mass hadrons, scalar quarkonium, tetra-quark 
states~\cite{jaffe} and glueballs are expected to be all mixed. 
How this can happen has been studied in certain simple models, 
see e.g.~\cite{4QT} and references therein. It is an issue to be explored how 
the presence of the tetra-quark modifies the EoS. It is worth noting that
using a toy model for constituent quarks and gluons implementing chiral and
scale symmetry breaking
a large $m_\sigma \sim 1$ GeV in matter-free space is consistent with the 
lattice result regarding the thermal behavior of the gluon condensate~\cite{cdm},
which is a conceivable scenario known from the vacuum phenomenology of the scalar 
mesons.

\section*{Acknowledgments}
I am grateful for fruitful collaboration with H.~K.~Lee, W.-G.~Paeng
and M.~Rho.
Partial support by the Hessian
LOEWE initiative through the Helmholtz International
Center for FAIR (HIC for FAIR) is acknowledged.


\end{document}